\documentclass[english,aps, prl, superscriptaddress, twocolumn,letterpaper,showpacs,notitlepage]{revtex4-1}
\usepackage[T1]{fontenc}
\usepackage[latin9]{inputenc}
\usepackage{color}
\usepackage{float}
\usepackage{bm}
\usepackage{amsmath}
\usepackage{amssymb}
\usepackage{graphicx}

\makeatletter

\providecommand{\tabularnewline}{\\}

\usepackage{amsbsy,amssymb,amsmath,bm,bbold}
\usepackage{mathptmx,newtxtext,newtxmath,xspace}
\usepackage{graphicx,color,epsfig,rotate}
\usepackage{fancyhdr}
\usepackage{gensymb}
\usepackage[colorlinks=true, 
                    linkcolor=blue, 
                    urlcolor=blue,
                    citecolor=blue]{hyperref}
\allowdisplaybreaks[2]

\usepackage[normalem]{ulem}
\newcommand{\stkout}[1]{\ifmmode\text{\sout{\ensuremath{#1}}}\else\sout{#1}\fi}

\definecolor{magenta}{cmyk}{ 0, 1, 0,0}

\makeatother

\usepackage{babel}
\begin{document}

\title{Dynamics and Instabilities of the Shastry-Sutherland Model}

\author{Zhentao Wang}

\affiliation{Department of Physics and Astronomy, The University of Tennessee,
Knoxville, Tennessee 37996, USA}

\author{Cristian D. Batista}

\affiliation{Department of Physics and Astronomy, The University of Tennessee,
Knoxville, Tennessee 37996, USA}

\affiliation{Quantum Condensed Matter Division and Shull-Wollan Center, Oak Ridge
National Laboratory, Oak Ridge, Tennessee 37831, USA}

\date{\today}
\begin{abstract}
We study the excitation spectrum in the dimer phase of the  Shastry-Sutherland model by using an unbiased variational method
that works in the thermodynamic limit. The method outputs dynamical correlation functions in all possible channels. 
This output is exploited to identify the order parameters with the highest susceptibility (single or multitriplon condensation in a specific channel)
upon approaching a quantum phase transition in the magnetic field versus the $J'/J$ phase diagram.
We find four different instabilities: antiferro spin nematic, plaquette spin nematic, stripe magnetic order,
and plaquette order, two of which have been reported in previous studies.
\end{abstract}

\pacs{~}

\maketitle
The Shastry-Sutherland model (SSM) has become a paradigmatic Hamiltonian of frustrated quantum
magnetism~\cite{Shastry1981,Miyahara2003} because it includes an exactly
solvable ground state~\cite{Shastry1981}, very heavy low-energy
excitations~\cite{Miyahara1999,Weihong1999,Fukumoto2000_2, Knetter2000_0,Knetter2000,Totsuka2001},
exotic phases obtained upon varying the ratio $J'/J$ between two competing exchange constants~\cite{Albrecht1996,Weihong1999,Koga2000,Koga2000_2,Knetter2000,Weihong2001,Lauchli2002,Munehisa2003,Lou2012,Corboz2013},
and a series of magnetic field induced magnetization plateaux~\cite{Miyahara1999,Muller-Hartmann2000,Momoi2000_0,Momoi2000,Fukumoto2000_1,Fukumoto2001,Misguich2001,Miyahara2003_prb,Dorier2008,Abendschein2008,Isaev2009,Takigawa2010,Nemec2012,Corboz2014,Schneider2016,Takigawa_magnetization_bookchapter}.
Its realization in SrCu$_{2}$(BO$_{3}$)$_{2}$~\cite{Kageyama1999,Miyahara1999,Ueda1999}
enabled various experimental studies, including magnetization~\cite{Kageyama1999,Onizuka2000,Sebastian2008,Jaime2012,Takigawa2013,Matsuda2013,Haravifard2016},
specific heat~\cite{Tsujii2011}, inelastic neutron scattering (NS)~\cite{Kageyama2000,Cepas2001,Gaulin2004,Kakurai2005,Zayed2014,McClarty2017,Zayed2017},
far-infrared~\cite{Room2000}, electron spin resonance (ESR)~\cite{Nojiri1999,Nojiri2003},
Raman scattering~\cite{Lemments2000}, and nuclear magnetic resonance
(NMR)~\cite{Kodama2002,Kodama2005,Takigawa2013}. These studies revealed that a 
finite Dzyaloshinskii-Moriya (DM) interaction~\cite{Dzyaloshinsky1958,Moriya1960}
must be added to the SSM in order to account for several properties of SrCu$_{2}$(BO$_{3}$)$_{2}$~\cite{Nojiri1999,Cepas2001,Nojiri2003,Gaulin2004,Miyahara2004,Kodama2005,Kakurai2005,Shawish2005,Cheng2007,Romhanyi2011,Romhanyi2015,McClarty2017}.

Despite the great theoretical efforts devoted to the SSM, the problem is still far
from being solved. Perturbative approaches are only applicable in  narrow regimes and
conventional numerical methods suffer from severe
size effects. {As a consequence,  the nature of the quantum phase diagram  
has been debated for a long time~\cite{Albrecht1996,Weihong1999,Koga2000,Koga2000_2,Knetter2000,Weihong2001,Lauchli2002,Munehisa2003,Lou2012,Corboz2013}.
It is thus desirable to develop and apply alternative approaches. The infinite projected entangled-pair states (iPEPS) is an example of an alternative approach that works in the thermodynamic limit~\cite{Jordan2008,Corboz2013,Corboz2014}. However, it relies heavily on the initial guess of the physical states
and it is difficult to extract dynamical responses.

In this Letter, we introduce an {\it unbiased} variational method  to calculate the excitation spectrum
and dynamical responses (susceptibilities) of the SSM in the dimer phase~\footnote{A numerical method is said to be ``variational'' when it satisfies the variational principle. Other examples of unbiased variational methods include the density matrix renormalization group~\cite{White1992,White1993} and matrix product method~\cite{Ostlund1995,Rommer1997}, etc.}. The method 
works in the thermodynamic limit and it complements alternative approaches like iPEPS.  The basic idea  was originally introduced to compute the single-hole 
dispersion of the square lattice $t$-$J$ model~\cite{Trugman1988,Trugman1990}. The same
method was  applied to the Shastry-Sutherland lattice $t$-$J$ model~\cite{Shawish2006,Haravifard2006}.
Here we exploit this idea to compute dynamical correlators and dominant instabilities.  By working in a reduced Hilbert space, {\it which
preserves all model symmetries}, we obtain low
energy excitations classified by  quantum numbers. We then predict the character of the neighboring phases by detecting the order parameter with highest susceptibility.
Besides confirming the previously reported plaquette order and antiferro spin-nematic phases,  we  find two new phases; namely,
a plaquette spin-nematic phase and stripe magnetic ordering,  induced by simultaneously  increasing the magnetic field and  $J'/J$. 
In particular, the plaquette spin-nematic phase explains the nature of the two-triplon states (pinwheels) that crystallize at higher field values~\cite{Corboz2014}.

We consider the spin-$\frac{1}{2}$ SSM under a magnetic field~\cite{Shastry1981}:
\begin{equation}
\mathcal{H}=J\sum_{\langle ij\rangle}\bm{S}_{i}\cdot\bm{S}_{j}+J^{\prime}\sum_{\langle\langle ij\rangle\rangle}\bm{S}_{i}\cdot\bm{S}_{j}-h\sum_{i}S_{i}^{z},\label{eq:SSM}
\end{equation}
where $\langle ij\rangle$ and $\langle\langle ij\rangle\rangle$
denote intradimer and interdimer neighbors. 
The unit cell has 4 sites (see Fig.~\ref{Fig:lattice}).
 The exact ground state for small enough $J^{\prime}/J$ and $h$ 
is a direct product of singlet states on all dimers~\cite{Shastry1981}.
The elementary excitations of this ``dimer phase'' are singlet-triplet excitations known as triplons.
Triplons are dressed by quantum fluctuations with a magnetic correlation length $\xi$ that increases with
$J'/J$. Methods that can account for the spatial extent of these quantum fluctuations should provide a good description of the low-energy excitation  spectrum of the dimer phase. 

\begin{figure}[tbp]
	\centering
	\includegraphics[width=1\columnwidth]{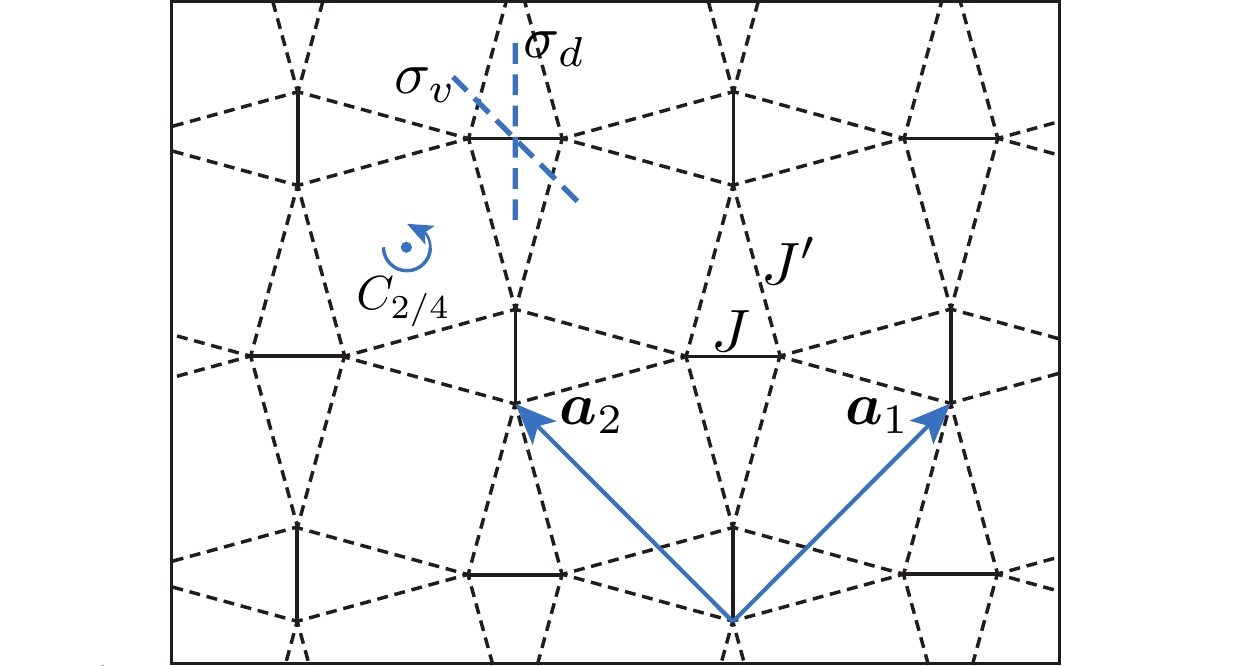}\caption{Lattice structure of the SSM. Intradimer and interdimer exchanges
		are denoted by $J$(solid line) and $J^{\prime}$(dashed line). The
		basis of the lattice is labeled by $\{\bm{a}_{1},\bm{a}_{2}\}$. The
		point group operations $\{\sigma_{v},\sigma_{d},C_{2},C_{4}\}$ are
		denoted accordingly.}
	\label{Fig:lattice}
\end{figure}


We start the process by creating local excited
states $|\varphi_{i}\rangle$ (e.g., single and two triplons). 
We then project these representative
states into subspaces with fixed momentum $\bm{k}$,
\begin{equation}
|\varphi_{i}(\bm{k})\rangle\equiv\frac{ \hat{P}_{\bm{k}} |\varphi_{i}\rangle}{\sqrt{\langle\varphi_{i}|\hat{P}_{\bm{k}}|\varphi_{i}\rangle}},
\end{equation}
where the projector is defined as $\hat{P}_{\bm{k}}\equiv\frac{1}{N}\sum_{\bm{r}}e^{i\bm{k}\cdot\bm{r}}\hat{T}(\bm{r})$.
$N\rightarrow\infty$ (thermodynamic limit) is the total number of unit cells, and $\hat{T}(\bm{r})$
is the translation operator. Application of $\mathcal{H}$
to  $|\varphi_{i}(\bm{k})\rangle$ generates
new states that dress the corresponding quasiparticle excitation. 
This procedure can be applied iteratively to systematically improve the variational space. 
After $M$ iterations, we obtain a basis $\{\varphi_{i}(\bm{k})\}$
with good quantum numbers $\bm{k}$ and $S_{\text{tot}}^{z}$. 
The number of iterations determines the spatial range $l$ of the fluctuations that dress the quasiparticle,
so the method is then expected to produce accurate results for $ l \gtrsim \xi $.

The  eigenvalues and eigenvectors of the Hamiltonian restricted to the variational space
are obtained by applying the implicitly restarted Arnoldi method~\cite{Lanczos1950,ARPACK}. 
The eigenvectors are classified by the Little Group of $C_{4v}$
for each momentum $\bm{k}$. {\it A continuous phase transition} manifests via a  
vanishing gap (condensation)  that  signals a phase transition  into 
a broken symmetry state. The symmetry of the new state is  determined by the irreducible
representation (IREP) of the eigenstate that becomes gapless. 
{\it To keep the method unbiased,   the
initial basis must not break the point group symmetry of $\mathcal{H}$}~\footnote{This means that the basis generates an invariant subspace of the group. In addition, a fully symmetric initial basis leads to better
convergence (as a function of $M$) in comparison with other choices~Ref.~\cite{Shawish2006}. }.

For illustration, we first focus on the $S_{\text{tot}}^{z}=0$ sector
relevant to $h=0$.  We include $\mathcal{D}=14$ $S_{\text{tot}}^{z}=0$
initial states to start the iteration~\cite{supp} and then
apply Eq.~(\ref{eq:SSM}) onto this basis to systematically increase the variational space~\footnote{The computational cost is exponential in the {\it linear} size of the quasiparticle. In the present case, the dimension of the truncated Hilbert space roughly scales as $\mathcal{D} \propto 6^M \propto 6^l$. For the $S_{\text{tot}}^z=0$ sector, the dimension of the variational space is $\mathcal{D}=74$ for $M=1$, $\mathcal{D}=396$ for $M=2$, $\ldots$, $\mathcal{D}=17\, 730\, 330$ for $M=8$.}. After obtaining the lowest energy eigenstates, we use the
eigenfunction to calculate $S_{\text{tot}}$ and its IREP~\footnote{In principle, one can switch the order of ``diagonalization'' and ``point group symmetry analysis,'' to reach larger $M$ and get better convergence. Similar tricks can be found in Refs.~\cite{Sinitsyn2007, Lauchli_ED_bookchapter}.}.

In contrast to the result obtained with perturbative continuous unitary transformations (CUTs)~\cite{Knetter2000},
we find that the first instability
as a function of $J^{\prime}/J$ (for $h=0$)  takes place in the $S_{\text{tot}}=0$ channel
with IREP $A_{2}$~\footnote{The $S=1$ instability found in Ref.~\cite{Knetter2000} is checked by our method both in $S_\text{tot}^z=0$ and $S_\text{tot}^z=1$ sectors. The energy difference between the two sectors is negligible.}. 
Figure~\ref{Fig2} shows the evolution of the gap as a function of $M$. 
Convergence is reached beyond $M=3$ for $J^{\prime}/J\lesssim0.5$, but the increase of $\xi$ slows down the convergence for 
larger  $J^{\prime}/J$.  Although Eq.~(\ref{eq:SSM}) does not conserve
the triplon number, the  state that condenses is adiabatically  connected
with the corresponding  $S_{\text{tot}}=0$ IREP $A_{2}$ pure two-triplon state in the $J^{\prime}/J \to 0$ limit (see Fig.~\ref{Fig2}).

\begin{figure}
\centering
\includegraphics[width=0.9\columnwidth]{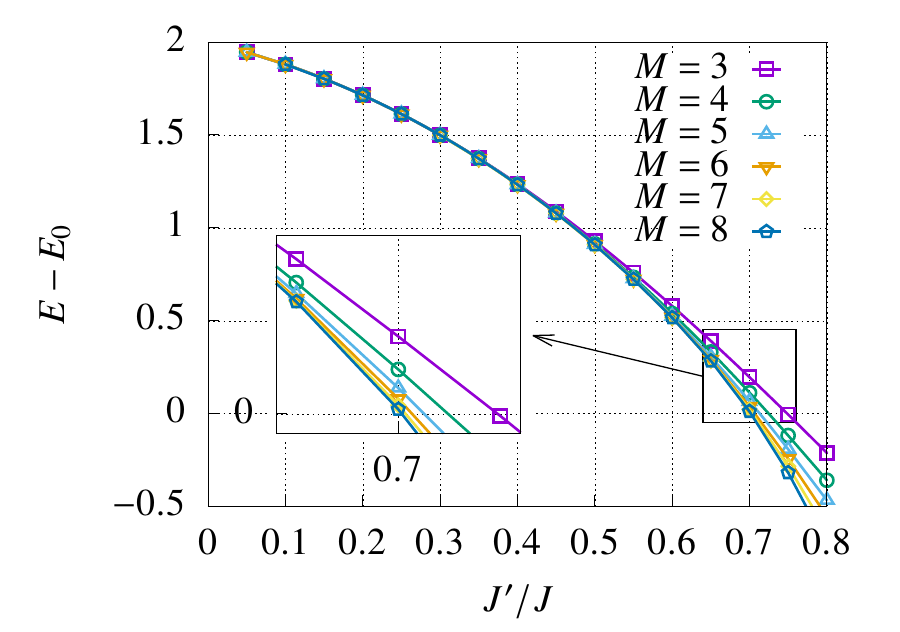}\caption{Gap of the lowest $S_{\text{tot}}=0$, $\bm{k}=(0,0)$ $A_{2}$
state, at $h=0$.}
\label{Fig2}
\end{figure}

We can read out the critical value of $J^{\prime}/J$ when this state
condenses. Figure~\ref{Fig3}(a) shows the evolution of the critical
value as we increase $M$ (circles). At $M=8$, $(J_c^{\prime}/J)^{(M=8)}\approx0.702$.
Previous tensor network based calculations~\cite{Lou2012,Corboz2013} showed that the
transition  is actually of first order and the transition point is  at $J_c^{\prime}/J=0.675$~\cite{Corboz2013}. A susceptibility analysis, like the one presented here, is in general inadequate to 
detect first order transitions. However, it can still be used to detect the nature of the order parameter {\it if the system still transitions into
the broken symmetry state with highest susceptibility}~\footnote{A typical example is the case of attractive interaction between the modes (particles) that become soft. Because of the attraction (negative quartic term in a Ginzburg-Landau expansion) the particle density changes discontinuously (first order transition) before the single-particle excitation becomes gapless. }. Given that the first-order transition takes place when this susceptibility is still finite, $J_c^{\prime}/J$ turns out to be smaller than the value at which the susceptibility becomes divergent}. This observation explains the difference between the values of $J_c^{\prime}/J$ obtained with both approaches. In addition, it  illustrates their complementary nature. The unbiased susceptibility analysis can be used to detect candidates for broken symmetry states. These candidates can then be tested with biased variational techniques, such as iPEPS, which can produce more accurate values of the transition point.

\begin{figure}
\centering
\includegraphics[width=1\columnwidth]{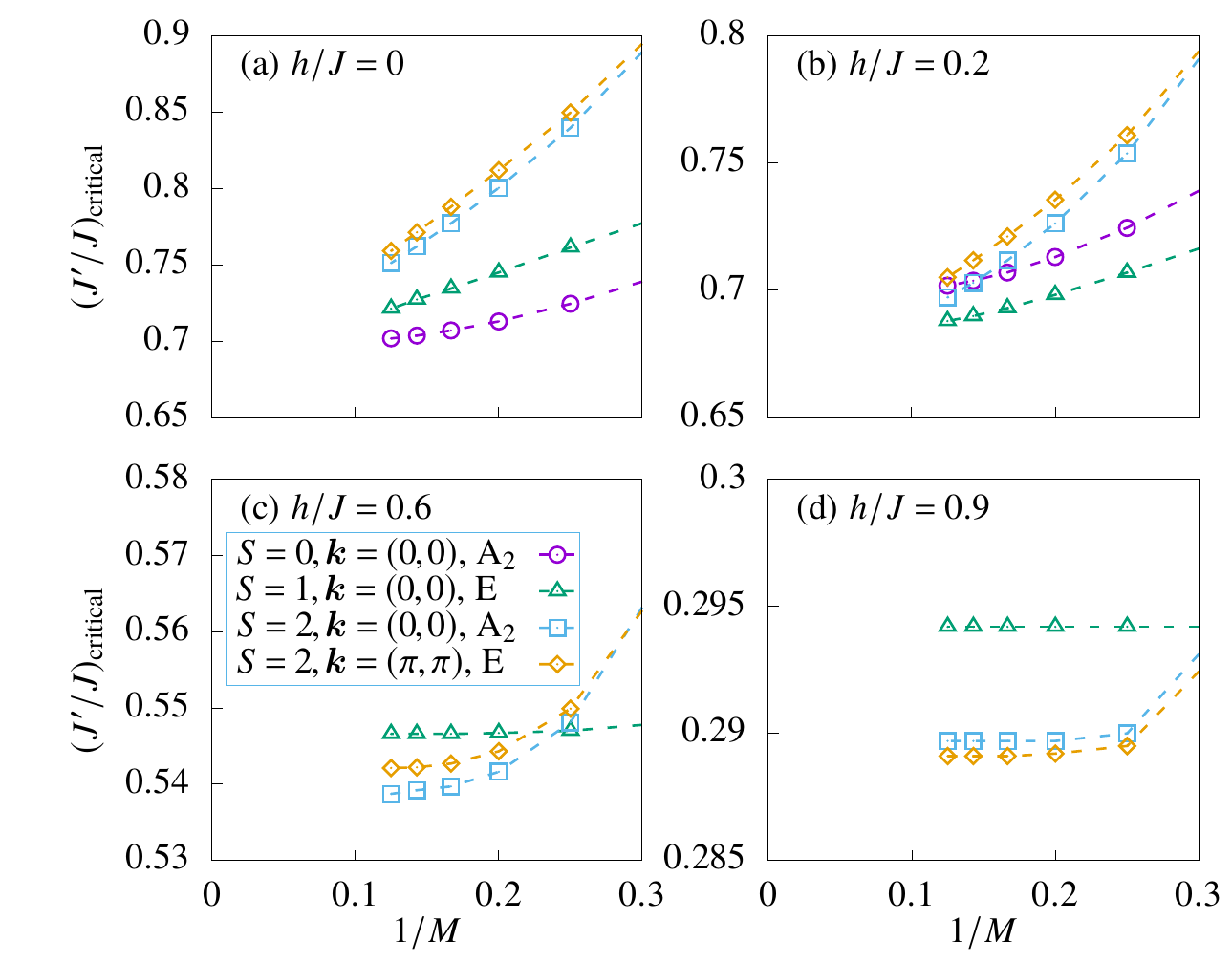}\caption{Critical values of $J^{\prime}/J$ for the condensation of different
states at 4 different magnetic fields. }
\label{Fig3}
\end{figure}

\begin{figure}
\centering
\includegraphics[width=1\columnwidth]{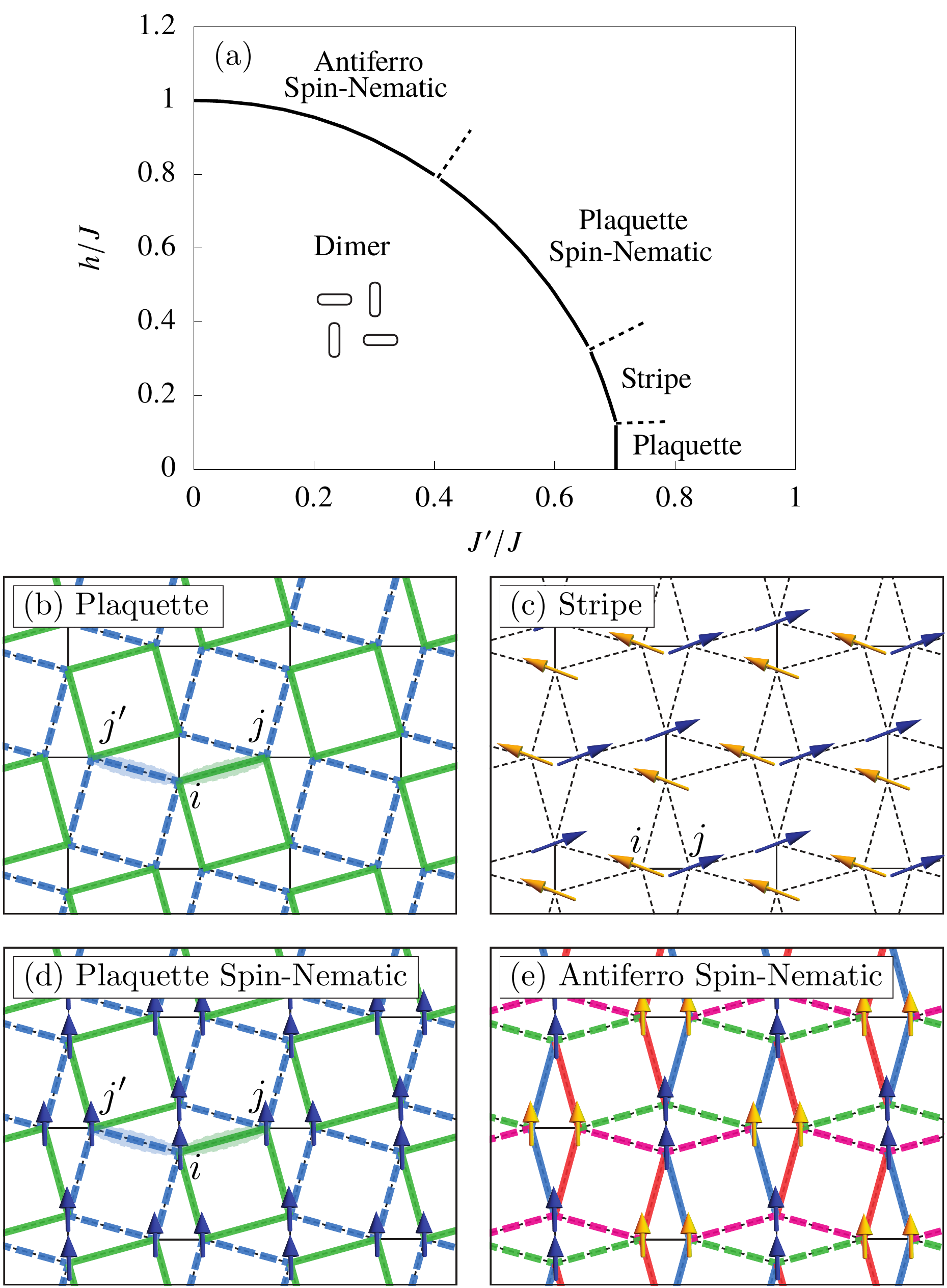}\caption{(a) Phase boundaries between the dimer state and its neighboring phases,
obtained from $M=8$ iterations. 
(b) Plaquette phase, order parameter defined as $\langle \bm{S}_i \cdot \bm{S}_j  - \bm{S}_i \cdot \bm{S}_{j^\prime}  \rangle$.
(c) Stripe phase, order parameter defined as $\langle \bm{S}_i - \bm{S}_j \rangle$.
(d) Plaquette spin-nematic phase, order parameter defined as $\langle S_i^+ S_j^+  - S_i^+ S_{j^\prime}^+  \rangle$.
(e) Antiferro spin-nematic phase (bond density wave), bonds with the same line (solid/dashed) but different colors have
opposite $\langle S_{i}^{+}S_{j}^{+}\rangle$ while bonds with different lines
have different $|\langle S_{i}^{+}S_{j}^{+}\rangle|$.}
\label{Fig4}
\end{figure}

Since the two-triplon bound state has $S_{\text{tot}}=0$, the
new ground state (condensate) must be nonmagnetic. Furthermore, since the $A_{2}$ state is odd (even)
under reflection (rotation)~\cite{supp}, the new ordered state should only break
reflection but not rotation symmetries. These features are consistent with the previously reported plaquette ordering~\cite{Koga2000,Lauchli2002,Lou2012,Corboz2013}.
 Figure~\ref{Fig4}(b) shows a schematic plot of the corresponding bond ordering.
As expected, $\langle\bm{S}_{i}\cdot\bm{S}_{j}\rangle$ becomes different 
on different plaquettes and there is no magnetic order. 
In other words, the plaquette order parameter can be defined as $\langle \bm{S}_i \cdot \bm{S}_j  - \bm{S}_i \cdot \bm{S}_{j^\prime}  \rangle$, where $ij$ and $ij^\prime$ are two bonds related by a mirror reflection [see Fig.~\ref{Fig4}(b)].

We consider now  the case of nonzero magnetic field. The energy of excited states with finite 
$S_{\text{tot}}^{z}$  decreases linearly in $h$.
Figure~\ref{Fig3}(b) shows that the dominant instability for  $h/J=0.2$ corresponds to 
condensation of a state  with $S_{\text{tot}}=1$, $\bm{k}=(0,0)$, and IREP $E$, leading to the stripe magnetic order depicted in 
Fig.~\ref{Fig4}(c). The IREP E is
a two-dimensional representation corresponding to the two possible  directions of the
stripes  (along $\bm{a}_{1}$ or $\bm{a}_{2}$). We note that
the two same-color spins in the same unit cell
are not identical (the corresponding mirror symmetry is broken). 

The stripe state has the highest susceptibility over a  narrow range  $0.66\lesssim J^{\prime}/J\lesssim0.70$
for $M=8$ iterations  [see Fig.~\ref{Fig4}(a)].  
Because of the frustrated exchange interactions, the energies of a few other states are not much higher than the stripe state~\cite{supp}.
Among them, the lowest one is a two-triplon state with $\bm{k}=(0,0)$ and IREP $B_1$, corresponding to vector chiral order~\cite{Chubukov2013, Parker2017, WangZT2017}.
Although their energies are slightly higher than the stripe magnetic instability within $M \leq 8$, the situation may change in the $M \rightarrow \infty$ limit, or if small perturbations are added to the original model. 

The $S_{\text{tot}}=2$ excited states take over for higher magnetic field values. 
Figure~\ref{Fig3}(c) shows that for $h/J=0.6$ the lowest excited state is the $S_{\text{tot}}=2$ two-triplon bound state with momentum $\bm{k}=(0,0)$ and IREP $A_2$.
The fact that this state and the $S_{\text{tot}}=0$ state that condenses at zero field  belong to the same point group  IREP $A_{2}$ indicates that the condensation
of the $S_{\text{tot}}=2$ $A_{2}$ state also leads to ``plaquette''
ordering [shown in Fig.~\ref{Fig4}(d)]; the difference being that the $S_{\text{tot}}=2$ condensate also  breaks the U(1) symmetry group of global spin rotations along the field direction, leading to spin-nematic ordering.
In other words, 
the local bond order parameter is $\langle S_{i}^{+}S_{j}^{+} - S_{i}^{+}S_{j^{\prime}}^{+}\rangle$ instead of 
$\langle\bm{S}_{i}\cdot\bm{S}_{j} - \bm{S}_{i}\cdot\bm{S}_{j'} \rangle$
($ij$ and $ij'$ denote two bonds connected by a mirror reflection $\sigma_d$, see Fig.~\ref{Fig4}).

As indicated in Fig.~\ref{Fig4}(a), the ``plaquette spin-nematic'' state covers a  wide range $0.40\lesssim J^{\prime}/J\lesssim0.66$. It has been shown in Ref.~\cite{Corboz2014} that
the $\frac{1}{8}$ plateau at slightly higher magnetic field values and
$J^{\prime}/J=0.63$ is induced by crystallization of $S_{\text{tot}}^{z}=2$
bound states. A closer scrutiny of the ``pinwheel'' structure of
these  bound states  shows that they
locally preserve rotational symmetries, while breaking reflection
symmetries~\cite{Corboz2014}; i.e.,  they are the same two-triplon  bound states that we are finding in the dilute limit. 

Moving away from the plaquette spin-nematic phase towards the $J^{\prime}/J \ll 1$ limit, 
it is already known from an early perturbative calculation that $S_{\text{tot}}=2$ two-triplon bound states with $\bm{k}=(\pi,\pi)$ give the highest susceptibility~\cite{Momoi2000}.
This is confirmed by our variational method [see Fig.~\ref{Fig3}(d)].
Since the two-triplon bound state has momentum $\bm{k}=(\pi,\pi)$, the corresponding ordered state also
breaks translational symmetry. As shown in Fig.~\ref{Fig4}(e), $\langle S_{i}^{+}S_{j}^{+}\rangle$
changes sign going from one unit cell to its nearest neighbors. 
Similar to the case of the  stripe ordering (which also comes from condensation of
IREP E states), there are two choices for aligning the bonds. 
We note that the breaking of the $C_4$ lattice rotational symmetry leads to a modulation of $\langle S_i^z \rangle$ that can be detected with NS experiments.

It is worth mentioning that the two spin-nematic states found in this Letter are different from nematic phases discussed in various other contexts~\cite{Fradkin2010,Kamiya11,Fernandes2012,Fernandes2014}.
The so-called ``Ising-nematic'' ordering corresponds to (discrete)  lattice rotation symmetry breaking. In contrast, ``spin-nematic'' ordering corresponds to  broken spin rotational symmetry.   The spin-nematic
orderings discussed in this Letter {\it break both the point group symmetry and
spin rotation symmetry}~\cite{Penc_nematic_bookchapter}.
In other words, they are {\it simultaneously} Ising nematic and spin nematic.

The frustrated nature of the SSM makes the calculation of dynamical response a difficult task. To date, the only calculation including multitriplon contributions is  the perturbative CUTs, which breaks down for $J^\prime/J \gtrsim 0.63$~\cite{Knetter2004}.
The variational Hilbert space generated by our method thus provides a more reliable access
to dynamical responses via the continued fraction method~\cite{Gagliano87}. 

\begin{figure}
\centering
\includegraphics[width=1\columnwidth]{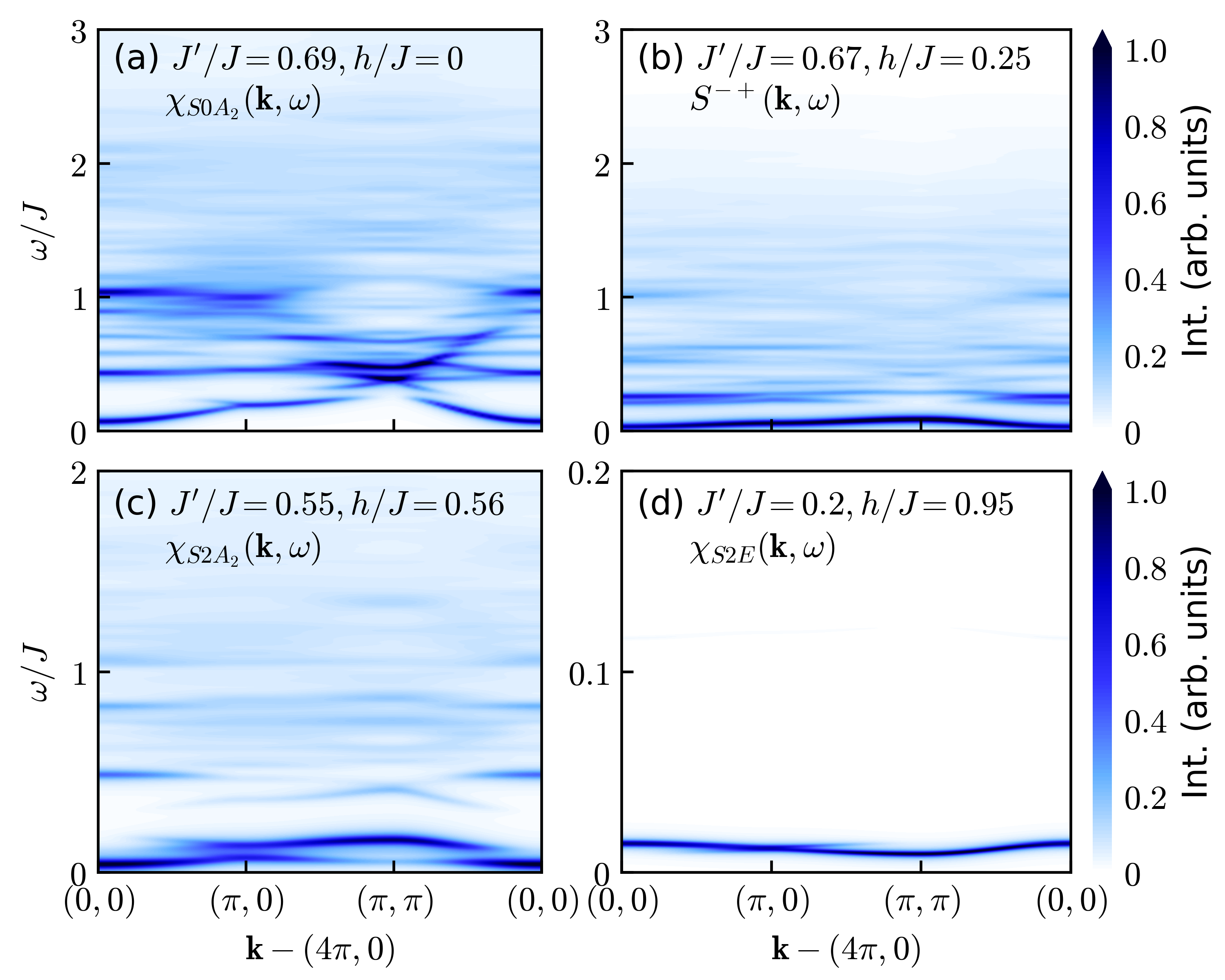}\caption{$T=0$ DSFs calculated near the phase boundaries at $M=8$. (a)-(c)
Lorentzian broadening factor $\eta=0.02J$ is used. (d) Lorentzian
broadening factor $\eta=0.001J$ is used.}
\label{Fig5}
\end{figure}

Near the phase boundaries, we expect the
susceptibilities of corresponding order parameters to diverge at $\omega=0$. 
Magnetic orderings, such as the stripe phase, are detected by computing the
dynamic structure factor (DSF)~\cite{fluc_diss,supp}:
\begin{equation}
S^{-+}(\bm{k},\omega)=2\pi\sum_{\nu}|\langle\nu|S_{\bm{k}}^{+}|0\rangle|^{2}\delta(\omega+E_{0}-E_{\nu}),\label{eq:sqw}
\end{equation}
which is measured with inelastic NS. As shown Fig.~\ref{Fig5}(b),
the lowest peak of $S^{-+}(\bm{k},\omega)$ approaches
$\omega=0$ near the phase boundary indicating condensation
of an $S_{\text{tot}}=1$ state.

The  divergent susceptibilities of the other phases are revealed by computing two-point dynamical 
correlation functions of the corresponding order parameters. These order parameters are the operators that
create a state that has finite overlap with the one that is condensing. For $J^{\prime}/J\ll1$, the lowest energy $S_{\text{tot}}=2$ eigenstates are known to be a linear
combination of triplons located on nearest (and next-nearest) neighbors~\cite{Momoi2000,supp}.
Denoting the order parameter as $A_{\bm{k}}^{S2E}$, 
the corresponding susceptibility is 
\begin{equation}
\chi_{S2E}(\bm{k},\omega)=2\pi\sum_{\nu}|\langle\nu|A_{\bm{k}}^{S2E}|0\rangle|^{2}\delta(\omega+E_{0}-E_{\nu}).
\end{equation}
Similarly, using the approximate wave functions of the $S_{\text{tot}}=2$
$A_{2}$ state and the $S_{\text{tot}}=0$ $A_{2}$ state~\cite{supp},
we can also construct the order parameters and compute the corresponding susceptibilities $\chi_{S2A_{2}}(\bm{k},\omega)$
and $\chi_{S0A_{2}}(\bm{k},\omega)$.  Figure~\ref{Fig5} shows the nearly divergent susceptibilities in each channel by picking
appropriate Hamiltonian parameters near the phase boundaries.

While the tendency toward stripe ordering can be detected with  inelastic NS,
the experimental detection of the other  phases is nontrivial. Lattice distortions  induced by
the order parameter through magnetostriction  can provide indirect evidence if they are large enough to be 
detected~\cite{Chubukov2013,Parker2017,WangZT2017}.
Experimental knobs, such as pressure, doping, and magnetic field can drive the material into different instabilities~\cite{Haravifard2016, Zayed2017}. 
Thus, the method presented in this Letter provides valuable insight for revealing the nature of the new phases in such experiments.  However, the model relevant to SrCu$_2$(BO$_3$)$_2$ also includes DM interactions that  modify the single triplon dispersion and  can potentially change the phase diagram reported here. In addition, DM interactions reduce the spin rotational symmetry of the model, implying that they can change the nature of the order parameters.

We emphasize that the applicability of the method is not restricted to the SSM considered here.
The same method can be used to detect the instabilities of other quantum paramagnets~\cite{Zapf14}. 
Especially,  it is very difficult to enumerate all the possible instabilities  for highly frustrated systems. 
The low energy spectrum produced by our method provides a valuable educated guess for biased numerical approaches.
Given that the method works in the thermodynamic limit, it can also detect incommensurate instabilities, which cannot be handled by most numerical methods.

In summary, we have used an unbiased variational method  to study the excitation spectrum
of the SSM in the dimer phase. Several instabilities are found next
to the dimer phase corresponding to condensations of single-triplon or two-triplon bound states. Two
of the instabilities (antiferro spin nematic and plaquette) are known
from previous studies and the others (plaquette spin nematic and stripe)
are newly discovered in this Letter. The same method can  be used to compute relevant 
dynamical correlation functions. 

We thank S.~Haravifard, B.~Shastry, G. Ortiz, S.~Zhang, and H.~Suwa for helpful
discussions. Z.W. and C.D.B. are supported by funding from the Lincoln
Chair of Excellence in Physics. This work used the Extreme Science
and Engineering Discovery Environment (XSEDE)~\cite{Towns2014} through
allocation TG-DMR170029, which is supported by NSF Grant No. ACI-1548562.
This research used resources of the Compute and Data Environment for Science (CADES) at the Oak Ridge National Laboratory, which is supported by the Office of Science of the U.S. Department of Energy under Contract No. DE-AC05-00OR22725.

\bibliographystyle{apsrev4-1}
\bibliography{ref}

\begin{thebibliography}{94}
\bibitem [{flu()}]{fluc_diss}%
  \BibitemOpen
  \href@noop {} {}\bibinfo {note} {According to the fluctuation dissipation
  theorem, the $T=0$ dynamic structure factor is equal to the imaginary part of
  the dynamic susceptibility up to a factor of $-2$.}\BibitemShut {Stop}%
\end{thebibliography}

\appendix
\setcounter{figure}{0} 
\renewcommand{\thefigure}{S\arabic{figure}} 
\setcounter{equation}{0} 
\renewcommand{\theequation}{S\arabic{equation}}
\newpage

\begin{center}   
{\bf ---Supplemental Material---} 
\end{center}

\section{Character table of $C_{4v}$ group}

The point group of Shastry-Sutherland lattice is $C_{4v}$, with 5 classes of symmetry operations:
identity $E$, two rotations $\{C_{4},C_{2}\}$ and two reflections
$\{\sigma_{v},\sigma_{d}\}$ (see Table.~\ref{Table:c4v}). 
\begin{table}[H]
\centering

\begin{tabular}{|c|c|c|c|c|c|}
\hline 
 & $E$ & $C_{2}$ & $2C_{4}$ & $2\sigma_{v}$ & $2\sigma_{d}$\tabularnewline
\hline 
\hline 
$A_{1}\left[1\right]$ & 1 & 1 & 1 & 1 & 1\tabularnewline
\hline 
$A_{2}\left[xy(x^{2}-y^{2})\right]$ & 1 & 1 & 1 & -1 & -1\tabularnewline
\hline 
$B_{1}[x^{2}-y^{2}]$ & 1 & 1 & -1 & 1 & -1\tabularnewline
\hline 
$B_{2}[xy]$ & 1 & 1 & -1 & -1 & 1\tabularnewline
\hline 
$E[x,y]$ & 2 & -2 & 0 & 0 & 0\tabularnewline
\hline 
\end{tabular}\caption{Chatacter table of group $C_{4v}$~\cite{Elliott_book}.}
\label{Table:c4v}
\end{table}

\section{Choice of initial basis}

\begin{figure}[bp]
\centering
\includegraphics[width=0.95\columnwidth]{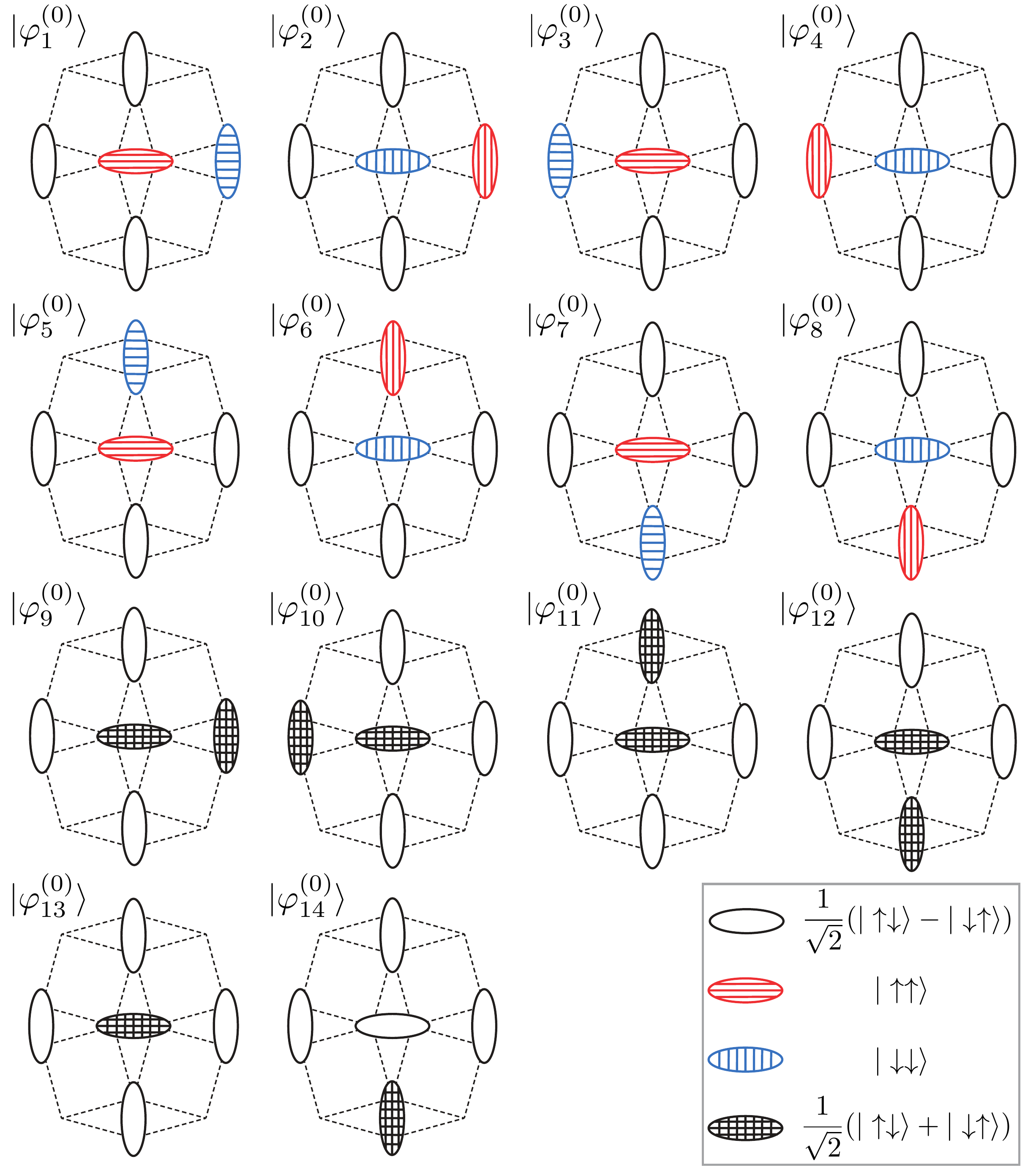}\caption{The $S_{\text{tot}}^{z}=0$ initial basis chosen in this Letter.}
\label{FigS1}
\end{figure}

\begin{figure}
\centering
\includegraphics[width=0.95\columnwidth]{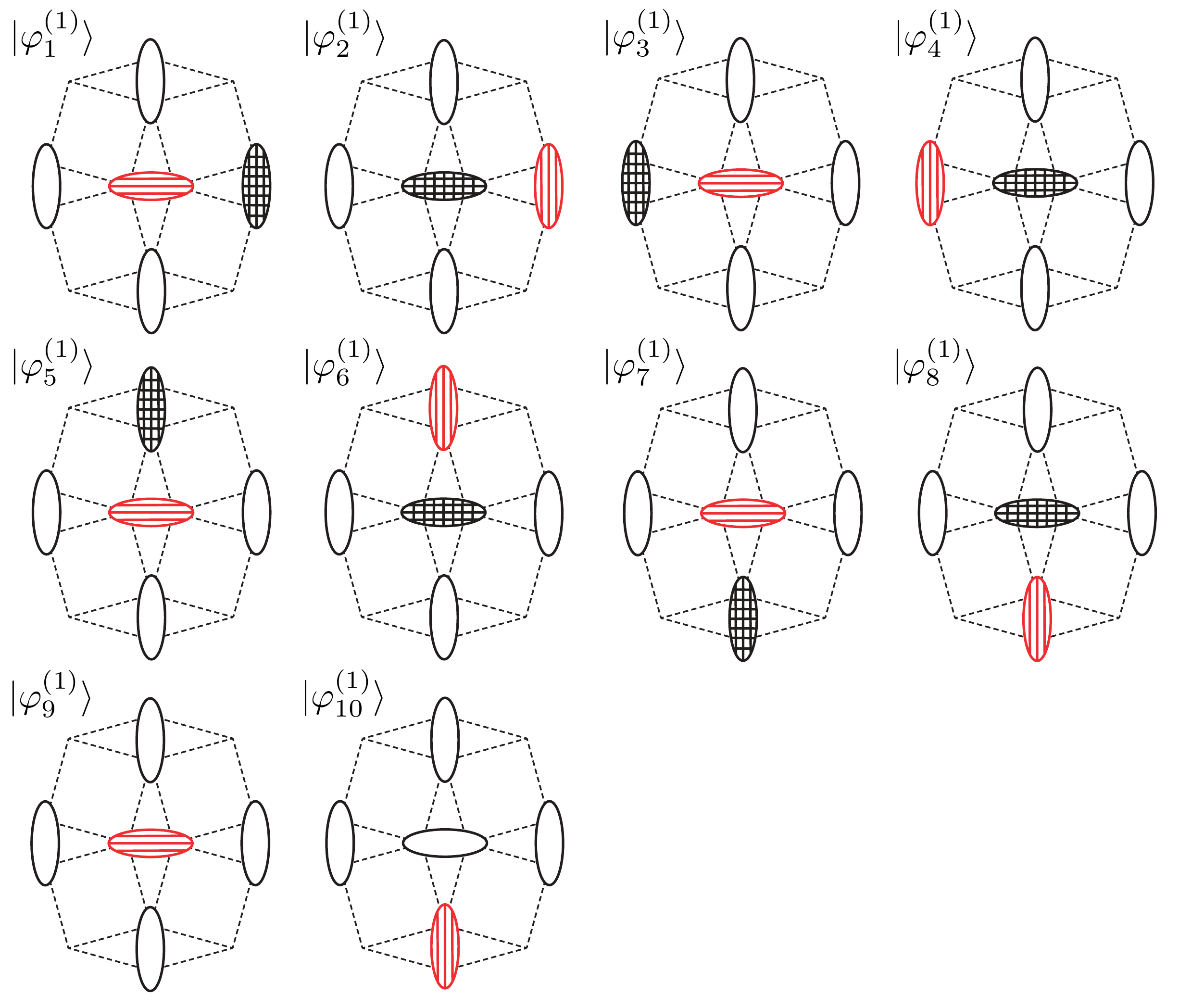}\caption{The $S_{\text{tot}}^{z}=1$ initial basis chosen in this Letter. The
notations are following Fig.~\ref{FigS1}.}
\label{FigS2}
\end{figure}

\begin{figure}
\centering
\includegraphics[width=0.95\columnwidth]{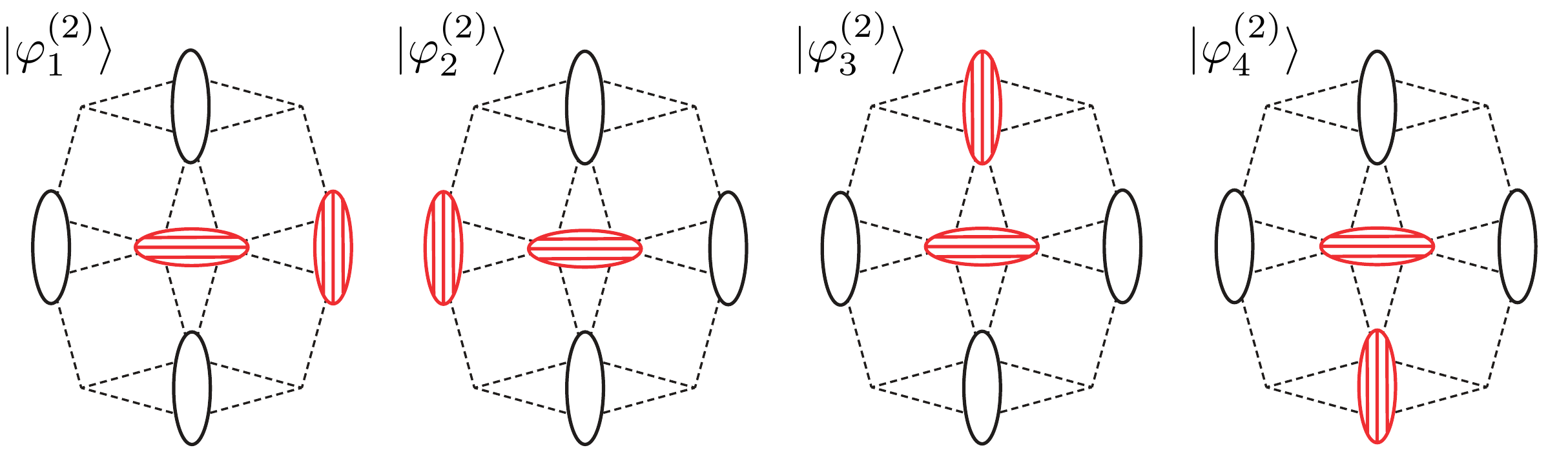}\caption{The $S_{\text{tot}}^{z}=2$ initial basis chosen in this Letter. The
notations are following Fig.~\ref{FigS1}.}
\label{FigS3}
\end{figure}

The initial basis is always chosen such that it does not break the
point group symmetries explicitly.
The choices for different values of $S_{\text{tot}}^{z}$ are listed in Figs.~\ref{FigS1},
\ref{FigS2} and \ref{FigS3}.

\section{Construction of order parameters}

\begin{figure}
\centering
\includegraphics[width=0.95\columnwidth]{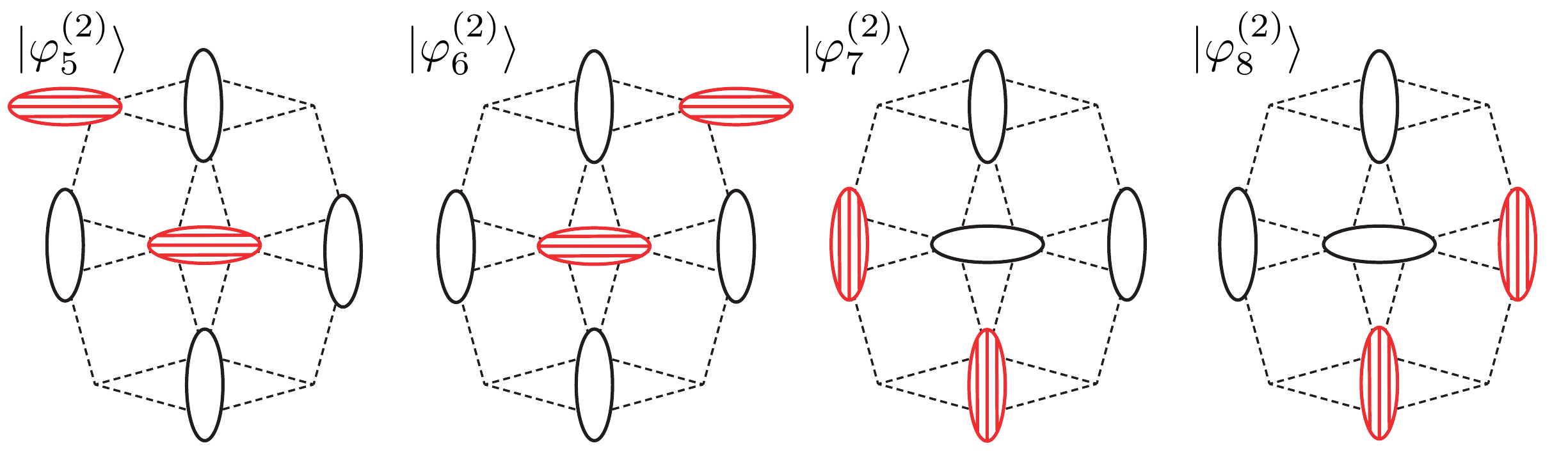}\caption{The extra $S_{\text{tot}}^{z}=2$ states for the approximate $S_{\text{tot}}=2$
bound state. The notations are following Fig.~\ref{FigS1}.}
\label{FigS4}
\end{figure}

In the $J^{\prime}/J\ll1$ limit, the $S_{\text{tot}}=2$, $\bm{k}=(\pi,\pi)$,
IREP $E$ states are known from Ref.~\cite{Momoi2000}.
Besides the $S_{\text{tot}}^{z}=2$ initial
states shown in Fig.~\ref{FigS3}, there is another set of relevant states
(see Fig.~\ref{FigS4}) generated at $M=2$. Here we write down the diagonalized wavefunctions
for $\bm{Q}\equiv (\pi,\pi)$:\begin{subequations}\label{eq:wf_S2E}
\begin{align}
 & \quad|\Psi_{1}(\bm{Q})\rangle_{S2E}\nonumber \\
 & \approx\alpha\left[|\varphi_{7}^{(2)}(\bm{Q})\rangle+|\varphi_{8}^{(2)}(\bm{Q})\rangle\right]\!+\!\beta\left[|\varphi_{2}^{(2)}(\bm{Q})\rangle-|\varphi_{1}^{(2)}(\bm{Q})\rangle\right],\\
 & \quad|\Psi_{2}(\bm{Q})\rangle_{S2E}\nonumber \\
 & \approx\alpha\left[|\varphi_{6}^{(2)}(\bm{Q})\rangle-|\varphi_{5}^{(2)}(\bm{Q})\rangle\right]\!+\!\beta\left[|\varphi_{4}^{(2)}(\bm{Q})\rangle-|\varphi_{3}^{(2)}(\bm{Q})\rangle\right],
\end{align}
\end{subequations}
where 
\begin{equation}
|\varphi_i^{(S_z)}(\bm{Q}) \rangle = \frac{\hat{P}_{\bm{Q}}|\varphi_i^{(S_z)} \rangle}{\sqrt{\langle \varphi_i^{(S_z)} | \hat{P}_{\bm{Q}} |\varphi_i^{(S_z)}  \rangle }}.
\end{equation}

The above wavefunctions have
$S_{\text{tot}}=2$, as long as the normalization condition $2|\alpha|^{2}+2|\beta|^{2}=1$
is satisfied. The values of $\alpha,\beta$ depend on 
$J^{\prime}/J$~\cite{Momoi2000}. To construct the nematic 
order parameter, we only need a bound state wavefunction 
that has significant overlap with the exact one.
We then choose $\alpha=0.7$, $\beta=0.1$ to
construct the order parameter 
\begin{equation}
A^{S2E}_{\bm{Q}}\equiv|\Psi_{1}(\bm{Q})\rangle_{S2E}\,\langle0|,\label{eq:Ak}.
\end{equation}
The momentum dependence of the corresponding susceptibility is obtained by computing the two-point correlator after replacing  $\bm{Q}$  with $\bm{k}$.


We have demonstrated in the main text
that the lowest bound state wave function has $S_{\text{tot}}=2$,
$\bm{k}=(0,0)$, IREP $A_{2}$ for larger values of $J^{\prime}/J$. Similarly, with the eight states in
Figs.~\ref{FigS3}-\ref{FigS4}, we can write down the approximate
$\bm{K}\equiv (0,0)$ wavefunction :
\begin{align}
 & \quad|\Psi(\bm{K})\rangle_{S2A_{2}}\nonumber \\
 & =\alpha\left[|\varphi_{5}^{(2)}(\bm{K})\rangle-|\varphi_{6}^{(2)}(\bm{K})\rangle-|\varphi_{7}^{(2)}(\bm{K})\rangle+|\varphi_{8}^{(2)}(\bm{K})\rangle\right]\nonumber \\
 & \quad+\beta\left[-|\varphi_{1}^{(2)}(\bm{K})\rangle-|\varphi_{2}^{(2)}(\bm{K})\rangle+|\varphi_{3}^{(2)}(\bm{K})\rangle+|\varphi_{4}^{(2)}(\bm{K})\rangle\right].\label{eq:wf_S2A2}
\end{align}

Once again, $S_{\text{tot}}=2$ is guaranteed by the normalization condition
$4|\alpha|^{2}+4|\beta|^{2}=1$. To construct the corresponding  order parameter 
$A^{S2A_2}_{\bm{k}}$, we use $\alpha=0.47$, $\beta=0.17$ in the main text.

We note that the exact bound state in this region  (larger $J^{\prime}/J$), 
has a larger size than the one in the $J^{\prime}/J \ll 1$ limit. Consequently, $A^{S2A_2}_{\bm{k}}$
produces more spectral weight at high energies in comparison to $A^{S2E}_{\bm{k}}$ [see Fig.~5(c)]. 

For even larger $J^{\prime}/J$ with $h=0$, the lowest states have $S_{\text{tot}}=0$,
$\bm{k}=(0,0)$, IREP $A_{2}$. 
we can similarly construct the approximate wavefunction at $\bm{K}\equiv (0,0)$:
\begin{align}
 & \quad|\Psi(\bm{K})\rangle_{S0A_{2}}\nonumber \\
 & =\alpha\Big[|\varphi_{1}^{(0)}(\bm{K})\rangle+|\varphi_{2}^{(0)}(\bm{K})\rangle+|\varphi_{3}^{(0)}(\bm{K})\rangle+|\varphi_{4}^{(0)}(\bm{K})\rangle\nonumber \\
 & \quad-|\varphi_{5}^{(0)}(\bm{K})\rangle-|\varphi_{6}^{(0)}(\bm{K})\rangle-|\varphi_{7}^{(0)}(\bm{K})\rangle-|\varphi_{8}^{(0)}(\bm{K})\rangle\Big]\nonumber \\
 & \quad-\beta\Big[|\varphi_{9}^{(0)}(\bm{K})\rangle+|\varphi_{10}^{(0)}(\bm{K})\rangle-|\varphi_{11}^{(0)}(\bm{K})\rangle-|\varphi_{12}^{(0)}(\bm{K})\rangle\Big].
\end{align}

In this case, the normalization condition $8|\alpha|^{2}+4|\beta|^{2}=1$
is different from the $S_{\text{tot}}=0$ condition, which requires
$\alpha=\beta$. Combining both conditions we have $\alpha=\beta=\frac{1}{2\sqrt{3}}$.

Once again, since this approximate wavefunction has a reduced overlap with the exact bound state, we see a significant amount of spectral weight at high energies in Fig.~5(a) in the main text.



\section{Dominant instability at zero field}
\begin{figure}
	\centering
	\includegraphics[width=0.8\columnwidth]{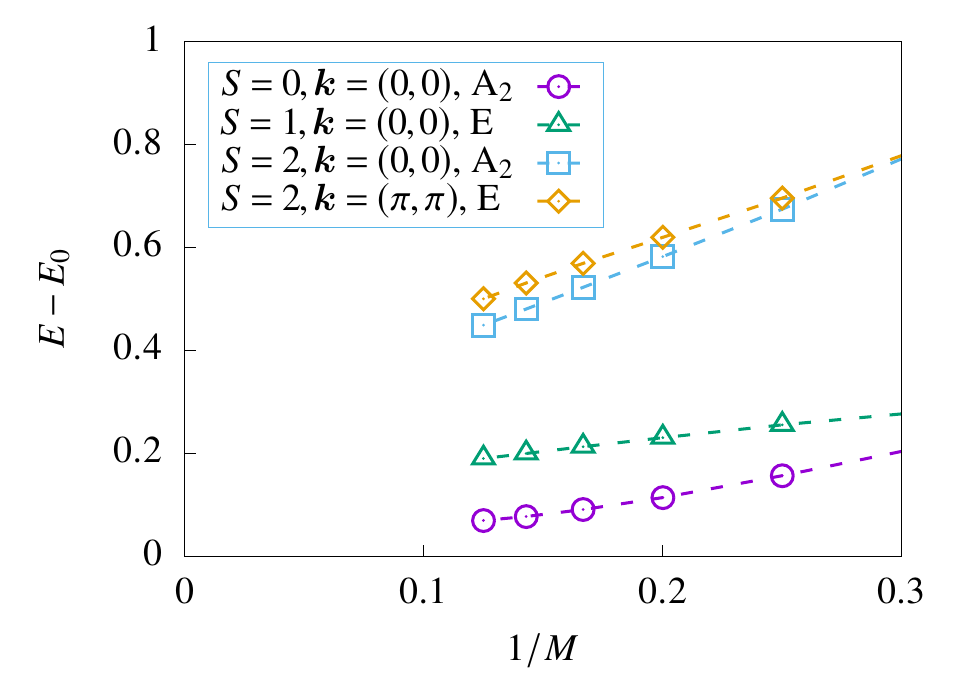}
	\caption{Gaps of the four candidate states at $J^\prime/J=0.69,\,h=0$. }
	\label{FigS5}
\end{figure}

In the $S_\text{tot}=0$ sector at $h=0$, we can roughly estimate $J_c^\prime/J \approx 0.69$ in the $M\rightarrow \infty$ limit (see Fig.~3 in the main text). At this point $J^\prime/J=0.69$, we show In Fig.~\ref{FigS5} that the gaps of the other $S_\text{tot}$ sectors remain finite in the $M\rightarrow \infty$ limit, even with a linear extrapolation. Note that the calculation converges for large enough $M$, implying that the slopes of the curves in Fig.~\ref{FigS5} should finally become zero. In other words, the true value of the gaps should be larger than that obtained from linear extrapolation.
Since all other gaps remain finite while the $S_\text{tot}=0$ gap is closing, we can safely conclude that the order parameter with highest susceptibility has $S_\text{tot}=0$ for $h=0$. 

\section{Convergence of dynamic structure factors}

Generally, the convergence speed of the DSFs is determined by $\xi$.
Larger $\xi$  values require larger number of iterations ($M$) to obtain qualitatively correct results. For instance, 
$\chi_{S2E}(\bm{k},\omega)$ is almost fully converged at $M=3$ and  $J^{\prime}/J=0.2$ [see Fig.~5(d) in the main text] 
because $\xi$ is  a few lattice spaces.

\begin{figure}
\centering
\includegraphics[width=1\columnwidth]{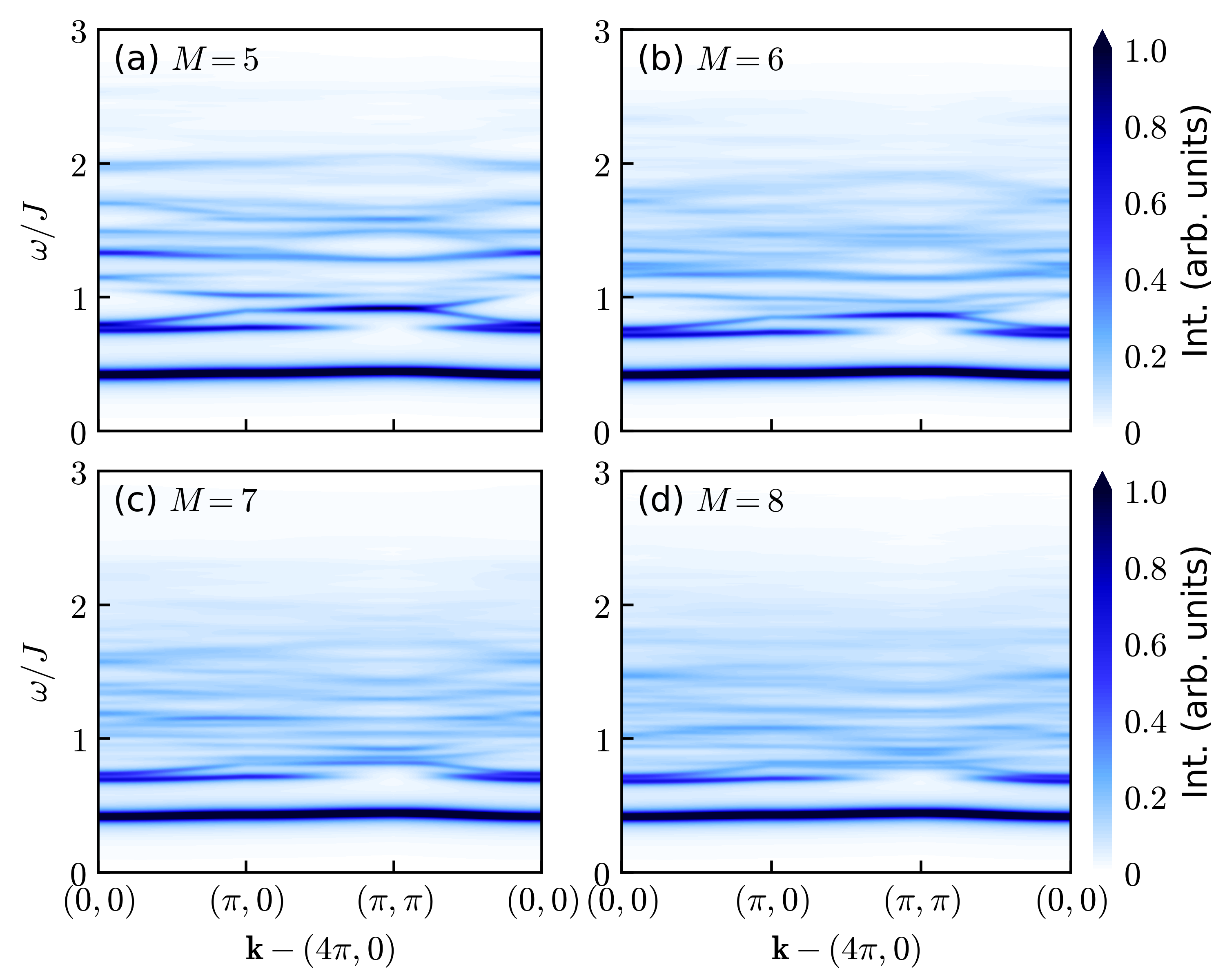}\caption{$T=0$ DSF $S^{-+}(\bm{k},\omega)$ for $J^{\prime}/J=0.63$, $h=0$,
at four different iterations $M=\{5,6,7,8\}$. Lorentzian broadening
factor $\eta=0.02J$ is used.}
\label{FigS6}
\end{figure}

\begin{figure}
	\centering
	\includegraphics[width=1\columnwidth]{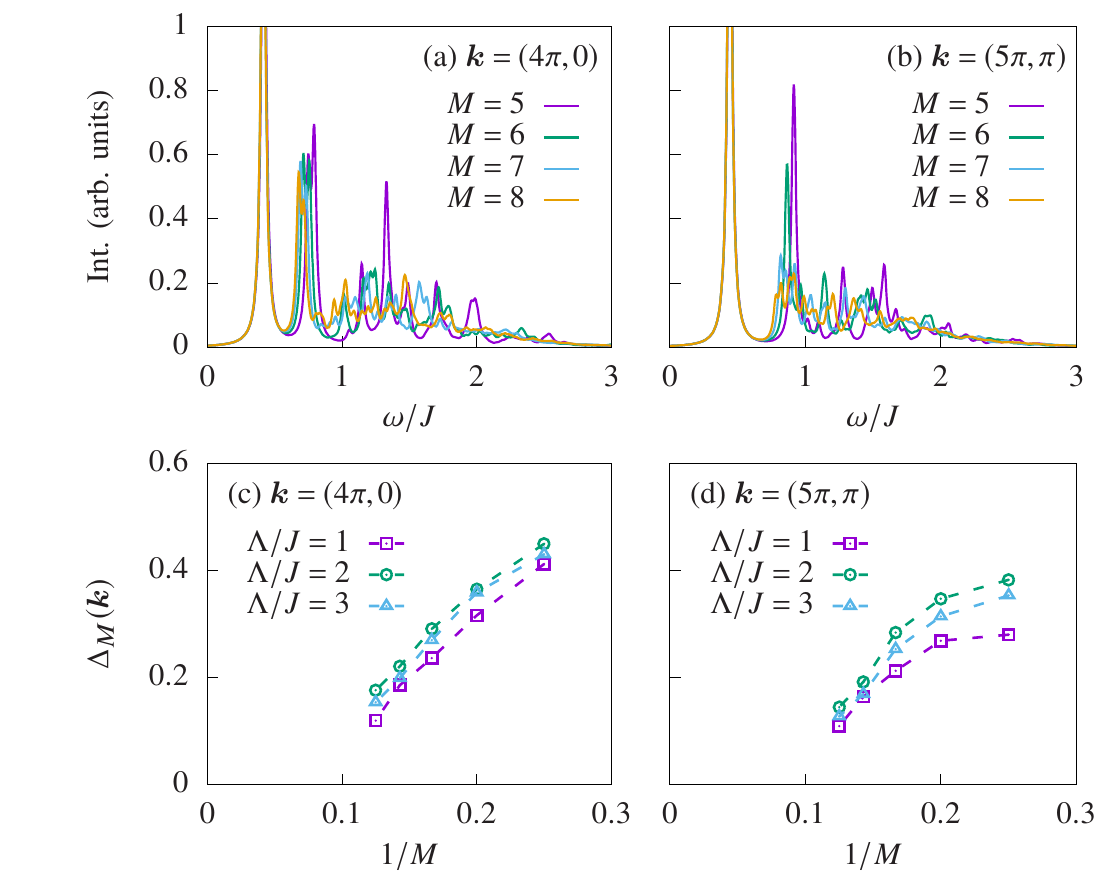}\caption{(a)(b) Linecuts of Fig.~\ref{FigS6} at two momenta. (c)(d) The relative error $\Delta_M(\bm{k})$ calculated at the two momenta, with three different choices of cutoff $\Lambda$.}
	\label{FigS7}
\end{figure}

To have a more quantative sense of the convergence as a function of
$M$, we first benchmark the DSF $S^{-+}(\bm{k},\omega)$ with $J^{\prime}/J=0.63$,
$h=0$, which is relevant for the compound SrCu$_{2}$(BO$_{3}$)$_{2}$~\cite{Corboz2014}.
In Fig.~\ref{FigS6}, we show the evolution of the calculated $S^{-+}(\bm{k},\omega)$
as a function of $M$. The lowest single and two-triplon
bands are already converged at $M=5$. We note that the method produces not only the bound states,
 but also the multitriplon continuum, which is qualitatively caputured at $M=5$ and about to
converge at  $M\gtrsim 8$ for the parameters used in Fig.~\ref{FigS6}.

To measure the relative change between different iterations, we can define
\begin{equation}
\Delta_M (\bm{k})= \sqrt{\frac{1}{\Lambda} \int_0^{\Lambda} d \omega \left( \frac{S^{-+}(\bm{k},\omega)_M - S^{-+}(\bm{k},\omega)_{M-1}}{S^{-+}(\bm{k},\omega)_M+ S^{-+}(\bm{k},\omega)_{M-1}} \right)^2},
\end{equation}
where $\Lambda$ is the energy cutoff. For the fully converged result, it is expected that $\Delta_M (\bm{k}) \rightarrow 0$.

In Fig.~\ref{FigS7}, we show the linecuts of Fig.~\ref{FigS6} at two momenta: $\bm{k}-(4\pi,0) =  (0,0) \text{ and } (\pi,\pi)$. Correspondingly, the relative errors $\Delta_M (\bm{k})$ are shown in the bottom panels of the same figure. The relative errors decrease rapidly as we increase $M$, reaching around $15\%$ at $M=8$. We note that the value of $\Delta_M (\bm{k})$ depends on the energy cutoff $\Lambda$: the lowest relative errors in Fig.~\ref{FigS7} are found when $\Lambda/J=1$, consistent with the fact that low-energy excitations converge faster than the high-energy ones. Note also that $\Delta_M (\bm{k})$ depends heavily on the choice of Lorentzian broadening factor used in the DSF calculation: larger broadening factors lead to smaller relative errors.

\begin{figure}[!htbp]
	\centering
	\includegraphics[width=1\columnwidth]{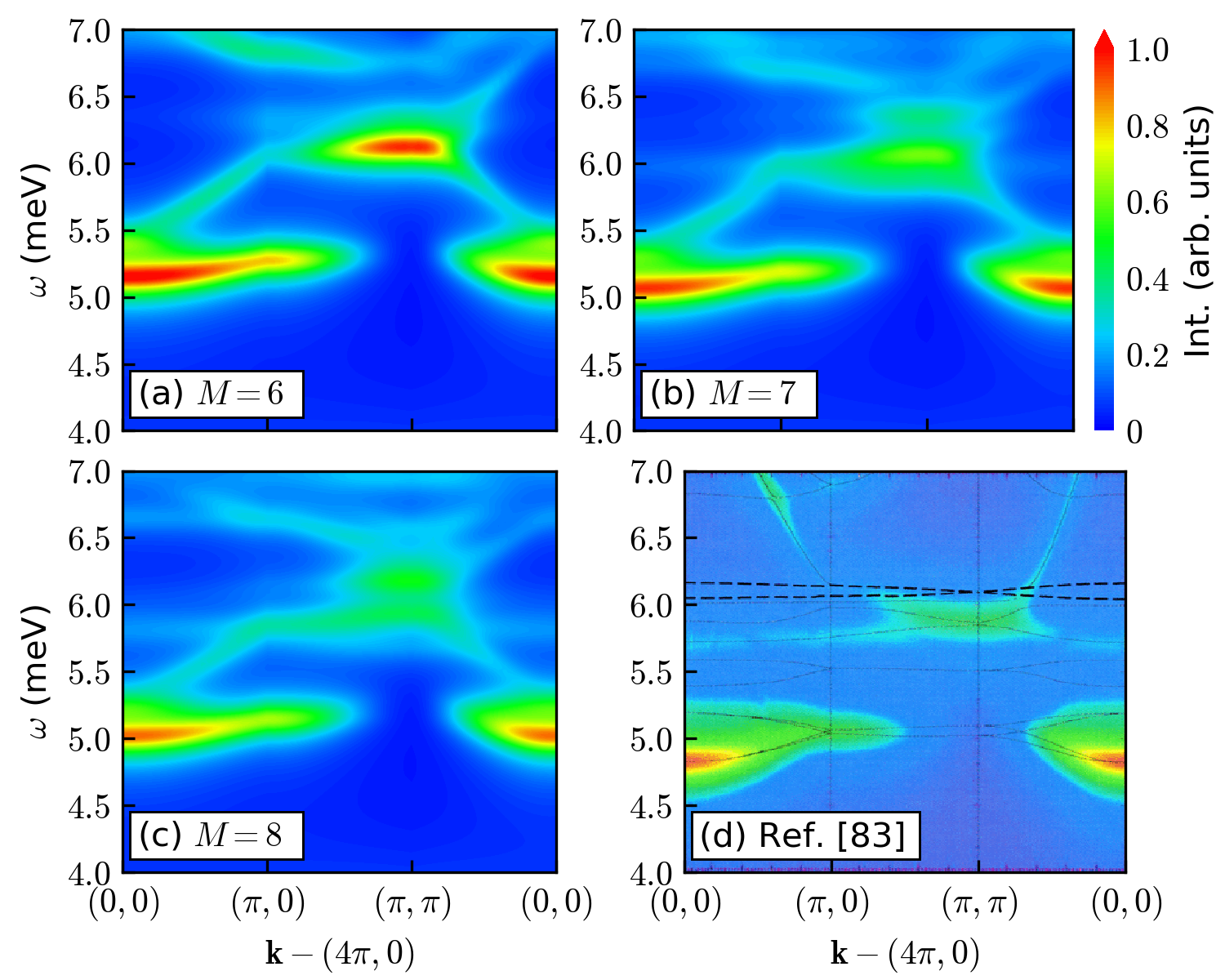}\caption{(a-c) $T=0$ DSF $S^{zz}(\bm{k},\omega)$ calculated using same parameters
		as Ref.~\cite{Knetter2004}: $J=6.16$meV, $J^{\prime}/J=0.603$,
		$h=0$. Lorentzian broadening factor $\eta=0.02J$ is used. (d) $S^{zz}(\bm{k},\omega)$
		reproduced from Ref.~\cite{Knetter2004}.}
	\label{FigS8}
\end{figure}

It is also worth benchmarking with Ref.~\cite{Knetter2004}, which
carried out the DSF calculation for SSM using the perturbative CUTs. Figure~\ref{FigS8} shows a
good qualitative agreement. However, there are two small differences: the band near $5$meV produced
by our method is slightly higher than that obtained from CUT, which
could arise from incomplete convergence for $M\lesssim8$ in our
method or missing higher order corrections in CUT. We also note that our method captures
more accurately the states in the continuum.

\section{Competing chiral state}
\begin{figure}[tbp]
	\centering
	\includegraphics[width=0.95\columnwidth]{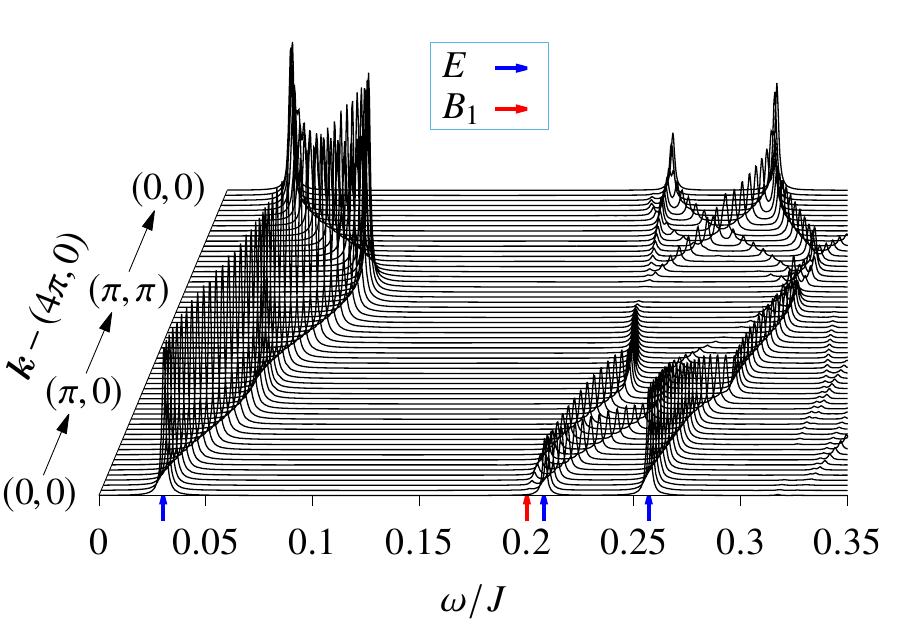}\caption{T=0 DSF $S^{-+}(\bm{k},\omega)$ calculated at $M=8$. The parameters
		are the same as in Fig.~5(b) in the main text: $J^{\prime}/J=0.67$,
		$h/J=0.25$. For better resolution, the Lorentzian broadening factor
		is chosen as $\eta=0.001J$ in this plot. The IREPs of the $\bm{k}=(4\pi,0)$
		peaks are labeled by the arrows.}
	\label{FigS9}
\end{figure}

The main text describes the $S_{\text{tot}}=1$ instability that  leads to the stripe magnetic order. 
As shown in Fig.~5(b),  the lowest instability from the DSF arises from 
condensation of the $\bm{k}=(0,0)$ mode. Figure~\ref{FigS9}
examines this point in more detail. Besides reproducing Fig.~5(b)
in higher resolution, we also denote the symmetries of the $\bm{k}=(4\pi,0)$
peaks {[}$\bm{k}=(0,0) \equiv (4\pi,0)${]}.

\begin{figure}[tbp]
	\centering
	\includegraphics[width=0.6\columnwidth]{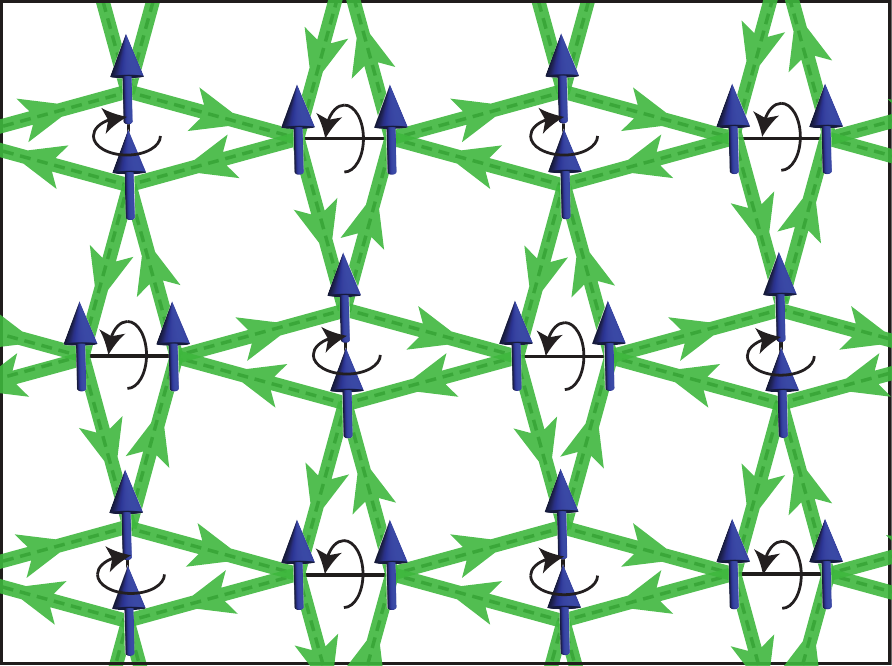}\caption{Schematic plot of the chiral phase. The bonds with arrow denote the
		sign of the vector chirality $\langle(\bm{S}_{i}\times\bm{S}_{j})\cdot\hat{z}\rangle$.}
	\label{FigS10}
\end{figure}

As we mentioned in the main text, the state that produces the lowest peak at $\bm{k}=(4\pi,0)$
belongs to the IREP $E$. In Fig.~\ref{FigS9}, we see that the next peak at $\bm{k}=(4\pi,0)$
corresponds to the IREP $B_{1}$ {[}with negligible weight at $\bm{k}=(4\pi,0)${]}.
Up to $M=8$, we find
that this $B_{1}$ peak is always higher in energy (lower susceptibility) than the one corresponding to the  IREP $E$.
It is worth mentioning that condensation of the  $B_{1}$ state leads to the  vector chiral ordering depicted in Fig.~\ref{FigS10}.

\end{document}